\documentclass{article}
\usepackage{spconf,amsmath,graphicx,multirow,amsfonts,amssymb}
\usepackage{float,hyperref}

\title{Investigating self-supervised learning for speech enhancement and separation}
\name{Zili Huang$^{1}$, Shinji Watanabe$^{2}$, Shu-wen Yang$^{3}$, Paola Garc\'ia$^{1}$, Sanjeev Khudanpur$^{1}$}
\address{
$^{1}$ Center for Language and Speech Processing and HLTCOE, Johns Hopkins University, USA \\
$^{2}$ Carnegie Mellon University, USA \\
$^{3}$ National Taiwan University, Taiwan \\
}
\begin{document}
\ninept

\maketitle
\begin{abstract}
Speech enhancement and separation are two fundamental tasks for robust speech processing. Speech enhancement suppresses background noise while speech separation extracts target speech from interfering speakers. Despite a great number of supervised learning-based enhancement and separation methods having been proposed and achieving good performance, studies on applying self-supervised learning (SSL) to enhancement and separation are limited. In this paper, we evaluate 13 SSL upstream methods on speech enhancement and separation downstream tasks. Our experimental results on Voicebank-DEMAND and Libri2Mix show that some SSL representations consistently outperform baseline features including the short-time Fourier transform (STFT) magnitude and log Mel filterbank (FBANK). Furthermore, we analyze the factors that make existing SSL frameworks difficult to apply to speech enhancement and separation and discuss the representation properties desired for both tasks. Our study is included as the official speech enhancement and separation downstreams for SUPERB.

\end{abstract}
\section{Introduction}
\label{sec:intro}
Speech enhancement and separation are two fundamental tasks for speech processing. The former suppresses background noises to improve speech quality and intelligibility while the latter extracts target speech from interfering speakers~\cite{wang2018supervised,michelsanti2021overview}. Both techniques are commonly used as preprocessing steps for tasks like automatic speech recognition (ASR) and speaker diarization, especially under noisy conditions~\cite{watanabe2020chime,medennikov2020stc,wang2021ustc}.

Over the past few years, deep learning-based methods have developed rapidly and become the mainstream for speech enhancement and separation, among which the supervised learning-based approaches are the most widely used ones. The supervised learning-based methods design objectives to approximate target signals either by estimating the spectral mask~\cite{weninger2015speech,soni2018time,fu2019metricgan,fu2021metricganplus,hershey2016deep,kolbaek2017multitalker,chen2017deep} or directly predicting the waveform~\cite{pascual2017segan,defossez2020real,luo2019conv,luo2020dual,chen2020dual,subakan2021attention,zeghidour2021wavesplit}.

Despite good performance, the supervised learning-based approaches are data-hungry and require a sufficient amount of labeled data to perform well, which is expensive. Self-supervised learning has been proposed to address this issue. Unlike supervised learning which directly optimizes for a specific task, SSL first pretrains models on unlabeled data to extract task-agnostic representations and then finetune models on the target domain. SSL has drawn massive attention due to its great performance and generalization ability. Inspired by the great success of SSL in natural language processing (NLP)~\cite{peters2018deep,devlin2019bert} and computer vision (CV)~\cite{misra2020self,he2020momentum}, an increasing number of SSL frameworks for speech have been proposed~\cite{schneider2019wav2vec,baevski2020wav2vec,hsu2021hubert} and successfully applied to various downstream tasks including ASR~\cite{baevski2020effectiveness}, speaker recognition~\cite{fan2020exploring}, emotion recognition~\cite{pepino2021emotion}, spoken language understanding~\cite{lai2021semi} etc.

To systematically explore the SSL paradigm for speech-related tasks, the Speech processing Universal PERformance Benchmark (SUPERB)~\cite{yang2021superb} is proposed. It evaluates the performance of a shared model across a wide range of speech processing tasks including phoneme recognition, ASR, speaker identification, automatic speaker verification, speaker diarization, intent classification, slot filling and emotion recognition, with minimal architecture changes and labeled data.

In this paper, we follow the principles of SUPERB~\cite{yang2021superb} and further investigate SSL for speech enhancement and separation. We aim to 1) Compare existing SSL models for speech enhancement and separation tasks. 2) Figure out the desired representation properties and proper pretraining setups for both tasks. We hope our study could cast light on the future design of SSL frameworks for speech enhancement and separation.

\section{Related Work}
\label{sec:related work}

\subsection{Speech separation}
With the advance of deep learning techniques, speech separation has witnessed rapid improvement. Most of the existing speech separation frameworks are based on supervised learning and can be divided into frequency-domain and time-domain methods. The former estimates time-frequency (T-F) mask for each source based on the STFT features and reconstructs individual sources using inverse short-time Fourier transform (iSTFT). Typical systems include Deep Clustering~\cite{hershey2016deep}, uPIT~\cite{kolbaek2017multitalker}, Deep Attractor Network~\cite{chen2017deep} etc. The time-domain methods~\cite{luo2019conv,luo2020dual,chen2020dual,subakan2021attention,zeghidour2021wavesplit} take the waveform of mixtures as input and directly predict the waveform of different sources using an encoder-decoder architecture. They are achieving state-of-the-art results in recent years. For SSL's applications in speech separation, in~\cite{huang2020self}, the authors find self-supervised pretraining on enhancement data can stabilize the label assignment during separation training and improve separation performance.

\subsection{Speech enhancement}
Over the past few years, deep learning-based enhancement models have dominated this field. Same as speech separation, speech enhancement methods can be divided into frequency and time domain. Among frequency-domain methods, ~\cite{weninger2015speech} uses a recurrent neural network (RNN) to predict T-F masks. MMSE-GAN~\cite{soni2018time} generates T-F masks using a generative adversarial network (GAN). MetricGAN~\cite{fu2019metricgan} and MetricGAN+~\cite{fu2021metricganplus} propose a method to train the generator with respect to enhancement evaluation metrics. Among time-domain methods, SEGAN~\cite{pascual2017segan} uses a GAN to directly generate the clean waveform. DEMUCS~\cite{defossez2020real} use an encoder-decoder architecture with skip-connections to predict the clean waveform. Unlike separation, where time-domain methods are dominating, frequency-domain methods are still competitive for speech enhancement~\cite{fu2021metricganplus}.

\section{Methodology}
\label{sec:methodology}
\subsection{Self-supervised pretrained models}
In this paper, we evaluate 13 SSL upstream models from the S3PRL toolkit~\cite{yang2021superb} on speech enhancement and separation downstream tasks. These SSL models can be categorized into generative and contrastive models~\cite{liu2021self}.

\textbf{Generative models} train an encoder to transform input $\mathbf{x}$ to representation $\mathbf{z}$, and try to reconstruct $\mathbf{x}$ with representation $\mathbf{z}$~\cite{liu2021self}. The generative models we studied include APC~\cite{chung2019unsupervised}, VQ-APC~\cite{chung2020vector}, NPC~\cite{liu2020non}, Mockingjay~\cite{liu2020mockingjay} and TERA~\cite{liu2021tera}. APC~\cite{chung2019unsupervised} follows a language model training style, and it uses a RNN to predict the future spectrum. VQ-APC~\cite{chung2020vector} adds a vector quantization (VQ) layer on top of APC model to better control the model capacity. NPC~\cite{liu2020non} is proposed as a non-autoregressive alternative to APC. It uses convolution architectures and predicts the center frame based on left and right context. Inspired by BERT~\cite{devlin2019bert}, Mockingjay~\cite{liu2020mockingjay} pretrains a Transformer encoder by predicting masked time frames. TERA~\cite{liu2021tera} extends Mockingjay by also predicting masked frequency bins.

\textbf{Contrastive models} also train an encoder to transform input $\mathbf{x}$ to representation $\mathbf{z}$ but to measure similarity~\cite{liu2021self}. Among the contrastive models we use, CPC~\cite{oord2018representation} combines predicting future observations with a contrastive loss InfoNCE. Modified CTC~\cite{riviere2020unsupervised} proposes several changes to the model architecture to improve training stability and model performance. wav2vec~\cite{schneider2019wav2vec} uses the same InfoNCE objective but a larger CNN architecture. vq-wav2vec~\cite{baevski2019vq} adds a VQ layer to wav2vec, enabling the direct use of NLP models like BERT on top of it. wav2vec 2.0~\cite{baevski2020wav2vec} incorporates the vq-wav2vec and BERT model into one end-to-end framework. Unlike BERT which predicts the masked tokens, wav2vec 2.0 still uses the contrastive loss as the objective. Inspired by DeepCluster~\cite{caron2018deep}, HuBERT~\cite{hsu2021hubert} performs offline clustering on representations, enabling it to avoid contrastive loss through directly predicting the cluster labels of the masked positions. UniSpeech-SAT~\cite{chen2021unispeech} and WavLM~\cite{chen2021wavlm} models are variants of the HuBERT model. The UniSpeech-SAT model adds an utterance-wise contrastive loss to enhance speaker information modeling and mixes original speech with interfering speakers as data augmentation. The WavLM model adds gated relative position bias to the Transformer structure and also uses utterance mixing augmentation (both interfering speech and noises are added).

In addition to these models, PASE+~\cite{ravanelli2020multi} borrows ideas from both generative and contrastive models. It performs multiple SSL tasks including feature generation and contrastive learning to learn robust speech representations.

\subsection{Downstream models for enhancement and separation}


\begin{figure}[!ht]
    \centering
    \includegraphics[width=1.0\linewidth]{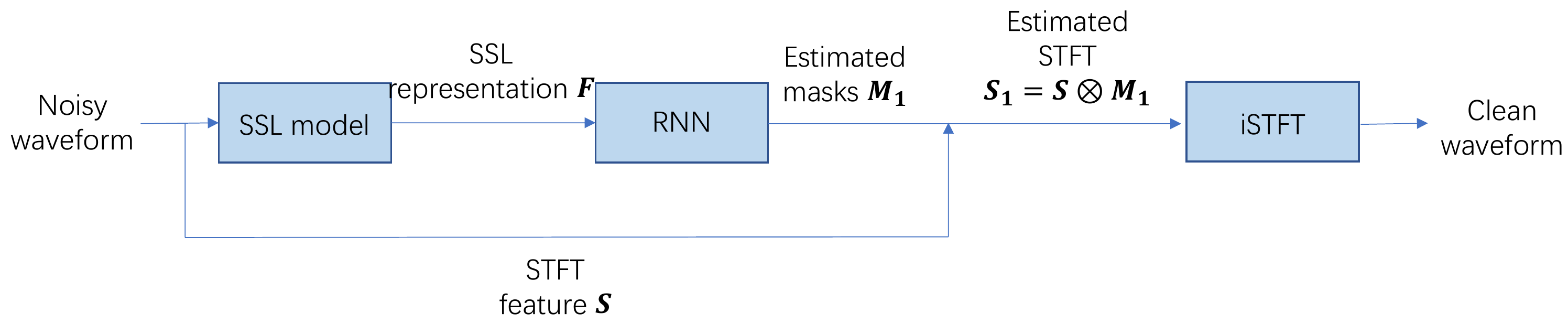}
    \caption{T-F mask-based speech enhancement downstream model. For speech separation, we estimate multiple masks from the RNN.}
    \label{fig:pipeline}
\end{figure}

Following the principles of SUPERB~\cite{yang2021superb}, we constrain our downstream models to be as lightweight as possible. After balancing between computational cost and performance, we choose a T-F mask-based model ~\cite{kolbaek2017multitalker} as our downstream model. As shown in Figure \ref{fig:pipeline}, for speech enhancement, the SSL model takes the noisy waveform as input and extracts speech representations $\mathbf{F}$. Based on $\mathbf{F}$, the RNN predicts the STFT mask $\mathbf{M_1}$ of the clean signal. The estimated mask $\mathbf{M_1}$ is multiplied with the STFT features $\mathbf{S}$ and transformed back to the time domain using iSTFT. The pipeline of separation is almost the same. The only difference is that the RNN will estimate multiple masks for different speakers. We use a three-layer bidirectional long short-term memory network (BLSTM) as the network architecture and the mean square error between the predicted mask and Ideal Non-negative Phase Sensitive Mask (INPSM)~\cite{kolbaek2017multitalker} is chosen as the objective. INPSM is defined as
$$M_s^{\mathrm{inpsm}} = \max\Big(0, \frac{|X_s(t, f)|\cos{(\theta_y(t,f)} - \theta_s(t,f))}{|Y(t, f)|}\Big)$$
where $Y$ is the mixture signal, $X_s$ is the signal from source $s$, $|X(t, f)|$ is the STFT magnitude of signal $X$ for time frame $t$ and frequency bin $f$, $\theta_y$ and $\theta_s$ are the phases of the mixture and source $s$.
For speech separation, permutation invariant training (PIT)~\cite{yu2017permutation} is utilized to address the speaker permutation problem. 

During finetuning, we follow SUPERB's setup~\cite{yang2021superb} to freeze the parameters of the SSL models. Instead of extracting representations from the last hidden layer, we weighted-sum the embeddings from all layers as the final representation $\mathbf{F}$ similar to ELMo~\cite{peters2018deep}. 
$$\mathbf{F} = \sum_{i=0}^{K-1} w_i \mathbf{F_i}$$
where K is the total number of layers, $\mathbf{F_i}$ is the representation extracted from the $i$th layer, $w_i$ is the weight for the $i$th layer. The layer weights $\mathbf{w} = [w_0, w_1, w_2, ..., w_{K-1}]$ are learned during the finetuning stage.

We did not use time-domain methods for the following two reasons. 1) \textbf{Stride difference}: Most of the existing SSL frameworks are using a stride size of 10 or 20ms, which corresponds to 160 to 320 samples for 16kHz audios. Such stride sizes are suitable for phoneme-level and sentence-level tasks such as ASR and speaker recognition but excessively large for time-domain speech enhancement and separation. As a comparison, most time-domain speech separation models are using a stride size smaller than 10 samples. We discuss the effect of stride size in Section \ref{sssec:stride}. 2) \textbf{Model complexity}: We could not find an appropriate time-domain method that is light enough for enhancement and separation. Putting a huge downstream model on top of SSL representations violates SSL's principle of simple finetuning.

\section{Experimental Setup}
\label{sec:setup}
\subsection{Dataset}
For speech enhancement, we use the Voicebank-DEMAND~\cite{valentini2017noisy}, a synthetic dataset created by mixing up clean speech and noise. The clean speech is extracted from the Voice Bank corpus~\cite{veaux2013voice}, and the noise is from the Diverse Environments Multichannel Acoustic Noise Database (DEMAND)~\cite{thiemann2013diverse}. The training set contains 28 speakers with 4 signal-to-noise ratios (SNRs) (15, 10, 5, and 0 dB) and the test set contains 2 speakers with 4 SNRs (17.5, 12.5, 7.5, and 2.5 dB). The training set contains 11,572 utterances (9.4h) and the test set contains 824 utterances (0.6h). The lengths of utterances range from 1.1s to 15.1s with an average of 2.9s.

For speech separation, we experiment on the LibriMix~\cite{cosentino2020librimix} dataset. The LibriMix dataset is simulated from the clean speech in LibriSpeech~\cite{panayotov2015librispeech} and noise in WHAM!~\cite{wichern2019wham}. Since most SSL models only support 16kHz audios as input, we choose the ``16kHz min" version of the data. The speech mixtures are created by mixing speech segments from different speakers. The loudness of each utterance is uniformly sampled between -25 and -33 loudness units relative to full scale (LUFS). Random noise samples with loudness between -38 and -30 LUFS are added to the speech mixtures. The training set contains 13,900 utterances with 43.3 hours of speech. In our experiments, we evaluate both ``sep\_clean" and ``sep\_noisy" conditions (separating speech from clean/noisy mixtures).

\subsection{Evaluation metric}
\label{sec:metric}
Speech enhancement requires both speech quality and intelligibility. In our experiment, we report two commonly used metrics: perceptual evaluation of speech quality (PESQ)~\cite{rix2001perceptual,itu862} and short-time objective intelligibility (STOI)~\cite{taal2010short}. PESQ measures the speech quality, and it predicts the subjective opinion scores of a degraded signal. We use the wide-band version of PESQ implemented in python-pesq~\cite{wang2019pesq}. STOI is a human-designed metric that shows a high correlation with the intelligibility of noisy speech. The range of STOI is from 0 to 100. For both metrics, a higher value indicates better performance.

For speech separation, we use Scale-Invariant Signal-to-Noise Ratio improvement (SI-SNRi) as the evaluation metric. It is a simpler and more robust alternative to Source-to-Distortion Ratio (SDR). SI-SNR is defined as
\begin{align*}
\mathbf{s}_{\mathrm{target}} &= \frac{(\mathbf{\hat{s}}^\intercal \mathbf{s})\mathbf{s}}{\left\| \mathbf{s} \right\|^2} \\
\mathbf{e}_{\mathrm{noise}} &= \mathbf{\hat{s}} - \mathbf{s}_{\mathrm{target}} \\
\text{SI-SNR}(\mathbf{s}, \hat{\mathbf{s}}) &= 10\log_{10}{\frac{\left\| \mathbf{s}_{\mathrm{target}} \right\|^2}{\left\| \mathbf{e}_{\mathrm{noise}} \right\|^2}}
\end{align*}
where $\mathbf{s} \in \mathbb{R}^L$ is the ground truth signal, $\hat{\mathbf{s}} \in \mathbb{R}^L$ is the estimated signal and $\left\| \mathbf{s} \right\| = \sqrt{\mathbf{s}^\intercal \mathbf{s}}$ denotes the $L^2$ norm of $\mathbf{s}$. SI-SNRi is the SI-SNR improvement against the mixtures, defined as
$$\text{SI-SNRi} = \text{SI-SNR}(\mathbf{s}, \hat{\mathbf{s}}) - \text{SI-SNR}(\mathbf{s}, \mathbf{m})$$
where $\mathbf{m} \in \mathbb{R}^L$ is the mixture signal.

\subsection{Model architecture and finetuning details}
During finetuning for speech enhancement and separation tasks, we use a three-layer BLSTM as the downstream model. Each BLSTM layer contains 896 hidden units. The output of the BLSTM is further processed by a linear layer and a ReLU activation.

The downstream models are finetuned for 150k steps with a batch size of 8. We use the Adam optimizer with a learning rate of $1e^{-4}$. Following SUPERB's~\cite{yang2021superb} setup, we don't decay the learning rate during finetuning. We choose the model with the best performance (highest PESQ for enhancement and SI-SNRi for separation) on the development set. 

\section{Experimental Results}
\label{sec:exp}

\subsection{Main experiment}
\label{ssec:main}
We present the speech enhancement and separation results for 13 SSL upstream models in Table \ref{tab:main}. For the STFT features, we use a frame size of 512, a frame shift of 160, and perform a 512-point FFT on each frame. The FBANK features are extracted using the torchaudio~\cite{yang2021torchaudio} toolkit with a frame size of 400 and a frame shift of 160. The number of Mel-frequency bins is set to 80. Delta and delta-delta coefficients are appended, and cepstral mean and variance normalization (CMVN) is applied. The extracted FBANK features have 240 dimensions. 

Among the SSL models, wav2vec2, HuBERT, UniSpeech-SAT, and WavLM use a stride of 320 samples (20ms) while other models use 160 samples (10ms). Among these SSL models, UniSpeech-SAT/WavLM Base+/Large, wav2vec2 Robust have seen noisy speech in real scenarios while other models are pretrained on the clean speech from audiobooks (LibriSpeech~\cite{panayotov2015librispeech} and LibriLight~\cite{kahn2020libri}). Our findings are as follows.

Compared with other tasks such as ASR, the improvement of SSL is not as large for enhancement and separation. For enhancement, only the HuBERT/UniSpeech-SAT/WavLM Large and UniSpeech-SAT Base+ achieve more than 0.05 PESQ improvement over the FBANK baseline. Other SSL models have comparable or even slightly worse performance. For separation, only the UniSpeech-SAT/WavLM Large can consistently outperform ($>$0.5dB SI-SNRi improvement) the STFT baseline for both sep\_clean and sep\_noisy conditions. The possible reasons for some SSL models don't perform well include 1) \textbf{Domain mismatch}. Most of the SSL models above are pretrained on the clean speech from audiobooks, and they have never seen noise and speaker overlaps before, making representations less robust to such conditions. For example, the Modified CPC and HuBERT Large achieve more than 0.5dB SI-SNRi improvement over the STFT baseline for the sep\_clean condition. However, their performance largely degrades when separating noisy mixtures. 2) \textbf{Information Loss}. The objectives of some SSL models encourage the systems to focus on global structures and build long-term dependencies. Some local information necessary for signal reconstruction is lost during pretraining.

Pretraining with audios from real scenarios seems to improve the enhancement and separation performance in some cases. The wav2vec2 Robust and WavLM Base+ largely improve the PESQ value for enhancement and slightly improve the SI-SNRi for separation. The UniSpeech-SAT Base+ performs almost the same as the UniSpeech-SAT Base for both tasks. The utterance mixing augmentation doesn’t seem useful for enhancement, but it improves the separation performance. Combining both techniques and other small modifications, the UniSpeech-SAT and WavLM Large models consistently outperform the HuBERT Large model. The UniSpeech-SAT Large model has achieved the best results for enhancement and separation tasks (except for the STOI metric). It improves the STFT and FBANK baselines by 0.15 PESQ, 0.8 STOI, 1.24/1.18 dB SI-SNRi on sep\_clean/sep\_noisy conditions.

Vector quantization seems to degrade the separation performance. VQ-APC and vq-wav2vec achieve worse separation performance compared to APC and wav2vec. A potential explanation is that converting continuous speech representations to discrete ones is detrimental to continuous sequence generation tasks like speech separation. Besides this, the TERA model improves both enhancement and separation performance over the Mockingjay, which shows that masked frequency bin prediction is useful for both tasks.

\begin{table}[ht!]

\centering
\caption{Evaluating 13 SSL upstream models on speech enhancement and separation downstream tasks. We measure speech enhancement performance with PESQ and STOI on the Voicebank-DEMAND~\cite{veaux2013voice} dataset. For speech separation, we evaluate on the Libri2Mix~\cite{cosentino2020librimix} dataset and report SI-SNRi for sep\_clean and sep\_noisy conditions.}
\footnotesize

\begin{tabular}{|l||c|c|c|c|}
\hline
\multirow{3}{*}{Model}  & \multicolumn{2}{c|}{Enhancement} & \multicolumn{2}{c|}{Separation} \\ \cline{2-5}
& PESQ$\uparrow$ & STOI$\uparrow$ & \multicolumn{2}{c|}{SI-SNRi (dB)$\uparrow$} \\ \cline{2-5}
& \multicolumn{2}{c|}{/} & sep\_c & sep\_n \\ \hline \hline

FBANK & 2.55 & 93.6 & 9.23 & 7.18 \\

STFT & 2.51 & 93.6 & 9.89 & 8.26 \\\hline

PASE+~\cite{ravanelli2020multi} & 2.56 & 93.9 & 9.87 & 8.01 \\\hline

APC~\cite{chung2019unsupervised} & 2.56 & 93.4 & 8.92 & 7.16 \\

VQ-APC~\cite{chung2020vector} & 2.56 & 93.4 & 8.44 & 6.86 \\

NPC~\cite{liu2020non} & 2.52 & 93.1 & 8.04 & 6.75 \\

Mockingjay~\cite{liu2020mockingjay} & 2.53 & 93.4 & 9.38 & 7.74 \\

TERA~\cite{liu2021tera} & 2.54 & 93.6 & 10.19 & 8.28 \\\hline

Modified CPC~\cite{riviere2020unsupervised} & 2.57 & 93.7 & 10.40 & 8.15 \\

wav2vec~\cite{schneider2019wav2vec} & 2.53 & 93.8 & 9.30 & 7.09 \\

vq-wav2vec~\cite{baevski2019vq} & 2.48 & 93.6 & 8.16 & 6.22 \\

wav2vec2 Base~\cite{baevski2020wav2vec} & 2.55 & 93.9 & 9.77 & 7.52 \\

wav2vec2 Large & 2.52 & 94.0 & 10.02 & 8.01 \\

wav2vec2 Robust~\cite{hsu2021robust} & 2.59 & 94.1 & 10.35 & 8.22 \\

HuBERT Base~\cite{hsu2021hubert} & 2.58 & 93.9 & 9.36 & 7.46 \\

HuBERT Large & 2.64 & 94.2 & 10.45 & 8.45 \\

UniSpeech-SAT Base~\cite{chen2021unispeech} & 2.60 & 94.0 & 10.33 & 8.28 \\

UniSpeech-SAT Base+ & 2.61 & 94.2 & 10.25 & 8.30 \\

UniSpeech-SAT Large & \textbf{2.70} & 94.4 & \textbf{11.13} & \textbf{9.44} \\

WavLM Base~\cite{chen2021wavlm} & 2.56 & 94.0 & 10.10 & 7.97 \\

WavLM Base+ & 2.60 & 94.0 & 10.58 & 8.68 \\

WavLM Large & 2.68 & \textbf{94.5} & 10.97 & 9.14 \\
\hline
\end{tabular}

\label{tab:main}
\end{table}

\subsection{Ablation studies}
\label{ssec:ablation}
In this section, we use the HuBERT model as an example to study the factors that influence the SSL model's performance on speech enhancement and separation tasks. 

\subsubsection{Effect of stride size}
\label{sssec:stride}
\begin{table}[t]
    \centering
    \caption{Speech enhancement and separation performance of the STFT and HuBERT Base/Large upstreams with different stride sizes. For separation, we only consider the sep\_clean condition.}
    \label{tab:stride}
    \begin{tabular}{c | c | c c | c} \hline
        Upstream & Stride & PESQ & STOI & SI-SNRi (dB) \\\hline\hline
        \multirow{2}*{STFT} & 160 & 2.51 & 93.6 & 9.89 \\
        & 320 & 2.42 & 93.3 & 8.79 \\\hline
        \multirow{2}*{HuBERT Base} & 160 & 2.68 & 94.1 & 10.47 \\
        & 320 & 2.58 & 93.9 & 9.36 \\\hline
        \multirow{2}*{HuBERT Large} & 160 & \textbf{2.80} & \textbf{94.5} & \textbf{11.26} \\
        & 320 & 2.64 & 94.2 & 10.45 \\\hline
    \end{tabular}
    \label{tab:stide}
\end{table}

As shown in Table \ref{tab:stride}, the stride size has a huge impact on speech enhancement and separation performance. For STFT, after we increase the stride size from 160 (10ms) to 320 (20ms), the PESQ, STOI, SI-SNRi (dB) degrade by 0.1, 0.3, and 1.1 respectively. The original stride of HuBERT Base/Large model is 320 (20ms). We upsample the representations by reducing the stride of the last convolution layer from 2 to 1. After upsampling, the HuBERT Base/Large models significantly outperform the original results. For all strides and metrics, the HuBERT models consistently outperform the STFT baseline. Note that even after upsampling, the stride size we use is still much larger than most time-domain enhancement and separation systems. As a comparison, we present the correlation between stride size and SI-SNRi for Conv-Tasnet~\cite{luo2019conv} in Table \ref{tab:conv-tasnet}. The vanilla Conv-Tasnet (with a stride of 8) achieves 14.34dB SI-SNRi on Libri2Mix. However, the performance degrades a lot as the stride size increases. When the stride size is larger than 160, the SI-SNRi of Conv-Tasnet is even lower than our STFT baseline. 

\begin{table}[t]
    \centering
    \caption{The separation performance of Conv-Tasnet~\cite{luo2019conv} on the 16kHz min Libri2Mix (sep\_clean condition) with different stride sizes. We use the Conv-Tasnet implementation from Asteroid~\cite{Pariente2020Asteroid} and adjust the stride size in the 1d convolution encoder}
    \label{tab:conv-tasnet}
    \begin{tabular}{c | c  c  c  c} \hline
        Stride & 8 & 40 & 160 & 320 \\\hline\hline
        SI-SNRi (dB) & 14.34 & 13.63 & 9.64 & 8.22 \\\hline
    \end{tabular}
    \label{tab:stide}
    \vspace{-3mm}
\end{table}

\subsubsection{Effect of layer weighting}
\label{sssec:layer}

For SSL models, different layers usually capture different speech information, which is used for different tasks. In this section, we extract speech representations from different layers of the HuBERT Large model and perform speech enhancement and separation on top of them. As shown in Table \ref{tab:layer}, the performance gap between different layers is significant. For speech enhancement, the embeddings from the 12th layer obtain the best PESQ and STOI numbers. It achieves around 0.1 PESQ and 0.6 STOI improvements compared to the last hidden layer. For speech separation, the performance declines as the layer becomes deeper, and the first layer outperforms the last layer by 4.21dB. The weighted-sum representations further improve the enhancement and separation results, and we observe that for most SSL models lower layers generally obtain higher weights. One possible explanation is that some local signal information necessary for speech reconstruction tasks is lost in deeper layers because it is restricted to local speech areas and less useful for objectives like contrastive learning and masked/future context prediction. Fully exploiting the information captured in different layers is important for speech enhancement and separation downstreams.

\begin{table}[H]
    \vspace{-3mm}
    \centering
    \caption{Speech enhancement and separation performance of different layer embeddings from the HuBERT Large model. For separation, we only consider the sep\_clean conition.}
    
    \begin{tabular}{c | c | c c | c} \hline
        Upstream & Layer & PESQ & STOI & SI-SNRi (dB) \\\hline\hline
        \multirow{4}*{HuBERT Large} & 0 & 2.52 & 93.9 & 9.96 \\
        & 12 & 2.58 & 94.0 & 8.58 \\
        & 24 & 2.49 & 93.4 & 5.75 \\
        & weighted & \textbf{2.64} & \textbf{94.2} & \textbf{10.45} \\\hline
    \end{tabular}
    \label{tab:layer}
\end{table}



\section{Conclusion}
\label{sec:conclusion}

In this paper, we investigate SSL for speech enhancement and separation. We evaluate 13 SSL upstream models on speech enhancement and separation with a T-F mask prediction downstream. Our experimental results reveal that 1) Although SSL models are not designed for waveform generation tasks like enhancement and separation, some of them achieve remarkable improvements over the STFT magnitudes and FBANKs. 2) Pretraining with audios from real scenarios and utterance mixing augmentation can increase the robustness of speech representations and improve the enhancement and separation performances. 3) Enhancement and separation require fine-grained waveform information to reconstruct the clean signal, which is often lost in deeper layers of SSL models. In the future, we will study SSL representations for time-domain methods.


\vfill\pagebreak

{
\scriptsize
\bibliographystyle{IEEEbib}
\bibliography{strings,refs}
}
\end{document}